\documentclass[aps,prb,reprint,amsmath,amssymb,superscriptaddress,floatfix]{revtex4-2}
\usepackage{graphicx,xcolor,hyperref}
\usepackage{braket}
\usepackage[ignoreunlbld,norefs,nocites]{refcheck}
\begin{document}

\title{Electronic-Entropy–Driven Solid-Solid Phase Transitions in Elemental Metals}
\author{S.\ Azadi}
\affiliation{Department of Physics and Astronomy, University of Manchester, Oxford Road, Manchester M13 9PL, UK}
\affiliation{Department of Physics, Clarendon Laboratory, University of Oxford, Parks Road, Oxford OX1 3PU, UK}
\email{sam.azadi@manchester.ac.uk}
\author{S.\ M.\ Vinko}
\affiliation{Department of Physics, Clarendon Laboratory, University of Oxford, Parks Road, Oxford OX1 3PU, UK}
\author{A.\ Principi}
\affiliation{Department of Physics and Astronomy, University of Manchester, Oxford Road, Manchester M13 9PL, UK}
\author{T.\ D.\ K\"{u}hne}
\affiliation{Center for Advanced Systems Understanding, Untermarkt 20, D-02826 G\"orlitz, Germany}
\affiliation{Helmholtz Zentrum Dresden-Rossendorf, Bautzner Landstra{\ss}e 400, D-01328 Dresden, Germany}
\affiliation{TU Dresden, Institute of Artificial Intelligence, Chair of Computational System Sciences, N\"othnitzer Stra{\ss}e 46 D-01187 Dresden, Germany}
\author{M.\ S.\ Bahramy}
\affiliation{Department of Physics and Astronomy, University of Manchester, Oxford Road, Manchester M13 9PL, UK}
\date{\today}

\begin{abstract}
We compute the thermodynamic phase diagram of seventeen elemental metals with hexagonal close-packed (hcp), face-centered cubic (fcc), and body-centered cubic (bcc) crystal structures using finite-temperature density functional theory. Helmholtz free-energy differences between competing hcp, fcc, and bcc phases are evaluated as functions of electronic temperature up to 7 eV, allowing us to identify solid–solid phase transitions driven by electronic entropy. The systems studied include Zr, Ti, Cd, Zn, Co, and Mg (hcp), Ni, Cu, Ag, Al, Pt, and Pb (fcc), and Cr, W, V, Nb, and Mo (bcc) in their ground-state structures. From the free-energy crossings, we extract the transition electronic temperatures and analyze systematic trends across the metallic systems. We found that all the studied systems go through one or two solid-solid phase transition caused purely by electronic entropy except Mg and Pb. Our results establish electronic entropy as a key factor governing structural stability in metals under strong electronic excitation.
\end{abstract}

\maketitle
\section{Introduction}
The response of transition metals to strong electronic excitation is governed by the intricate interplay between their partially filled d-bands, their lattice degrees of freedom, and the entropic contributions arising from thermally broadened electronic state occupations \cite{Grimvall,Porter,Kotani,Kirilyuk}. When electrons are driven far from equilibrium, for example under ultrafast laser irradiation, electrical breakdown, shock compression, or high-energy particle impact, the electronic subsystem can reach effective temperatures of several electron volts (eV) on femtosecond timescales, while the ions remain comparatively cold \cite{Recoules,Rousse,Medvedev,Sciaini,Amouretti2025,Celin2025,Kang2025,Mazevet, Humphries, Williamson, Silvestrelli,Beaurepaire,Hohlfeld,Zhang0,Carva,Kang2025}. Under these conditions, the Helmholtz free energy $F(T, V) = E(T, V) - T S(T, V)$, and in particular the electronic-entropy term $TS$, becomes the dominant thermodynamic quantity governing the relative stability of crystal phases, the forces on ions, and the effective elastic and vibrational properties.

The physical relevance of finite-temperature electronic effects in the ultrafast regime can be understood within the framework of the two-temperature model (TTM) \cite{Kaganov1957,Petrov2021,Allen1987,Carpene2006,Anisimov1975}, which separates the electron and lattice subsystems through their distinct thermalization timescales. In metals, electron–electron scattering occurs on the order of a few femtoseconds, rapidly establishing a Fermi–Dirac distribution characterized by an elevated electronic temperature T. In contrast, electron–phonon coupling acts on a substantially longer timescale, typically hundreds of femtoseconds to several picoseconds during which the lattice remains comparatively cold. Under such conditions, the system can transiently access electronic temperatures of several electronvolts while the ions remain near their initial positions, meaning the relevant thermodynamic potential governing structural stability is the electronic Helmholtz free energy $F(V,T)$ at fixed ionic configuration. Within this regime, the electronic-entropy-driven thermal pressure can cause a stabilization of a new phase before significant heat transfer from the electrons to the lattice. The resulting phase transformation is therefore a non-thermal, electronically driven solid–solid transition, governed by the electronic free-energy landscape rather than by lattice heating. Although we do not simulate real-time dynamics, the separation of timescales inherent to the TTM provides the physical justification for treating the transition as electronically controlled and for analyzing the phase stability solely in terms of the fixed-volume finite-temperature electronic free energy.

A further distinction between the electronically driven solid–solid transition and a conventional lattice-temperature–driven phase transformation lies in the characteristic timescales associated with the underlying atomic motions. A pressure-driven structural transformation responds to the instantaneous forces acting on the ions and therefore proceeds on timescales set by acoustic phonon propagation, meaning on the order of the sound velocity and the period of the lowest-frequency phonon modes. In most of crystalline transition metals, these timescales are tens to hundreds of femtoseconds, reflecting the speed at which ionic displacements respond to changes in the stress tensor. By contrast, a lattice-temperature–driven transition requires substantial energy transfer from electrons to ions, followed by anharmonic lattice motion and thermally activated barrier crossing, all of which occur on much longer timescales governed by electron–phonon coupling (hundreds of femtoseconds to picoseconds) and by the slower thermal diffusion of vibrational energy \cite{Baroni}. Consequently, once the electronic subsystem is rapidly heated to high T, the accompanying electronic-entropy  thermal pressure can drive a structural rearrangement on a timescale limited only by elastic response. 

Electronic entropy plays a central role in transition metals because their electronic structure near the Fermi level is highly non-uniform. The narrow d-band manifold produces large derivatives of the density of states (DOS) within a few tenths of an eV of the Fermi energy $E_F$, causing the chemical potential, internal energy, and electronic pressure to become strongly temperature dependent. As the electronic distribution broadens with increasing T, states deep within the d-bands, and in some cases states above the Fermi energy, become partially occupied. This reoccupation modifies screening, alters the Hartree and exchange–correlation potentials, modifies bonding character, and renormalizes the electronic contribution to elastic constants and phonons. These effects, proportional to energy integrals weighted by the entropy kernel $f(1-f)$ with $f$ being the Fermi-Dirac distribution function, can operate on energy scales of several eV and lead to qualitatively different material behavior compared to the T=0 ground state.

A proper theoretical treatment of these phenomena requires a finite-temperature electronic free-energy functional. Ground-state Kohn–Sham DFT accounts only for the T = 0 limit of the electronic degrees of freedom and is inadequate when the electronic subsystem is highly excited \cite{Hohenberg}. In contrast, Mermin’s generalization of density-functional theory provides a variational principle for the electronic grand potential at finite temperature and naturally incorporates Fermi–Dirac occupations, electronic entropy, and temperature-dependent screening in a fully self-consistent manner \cite{Mermin,Jones2014,Martin2004,Weinert92}. Within this formalism, both the internal energy $E(T)$ and the entropy $S(T)$ emerge from the same set of temperature-broadened Kohn–Sham states and the corresponding self-consistent potentials. As a consequence, finite-temperature DFT captures the reconstructed band structure, shifts in the chemical potential, enhancements of the electronic pressure, and entropy-driven free-energy stabilization or destabilization of crystal structures \cite{Kresse96,Zhang2021}.
\begin{figure*}[!htb]
    \centering
    \includegraphics[width=1.0\linewidth]{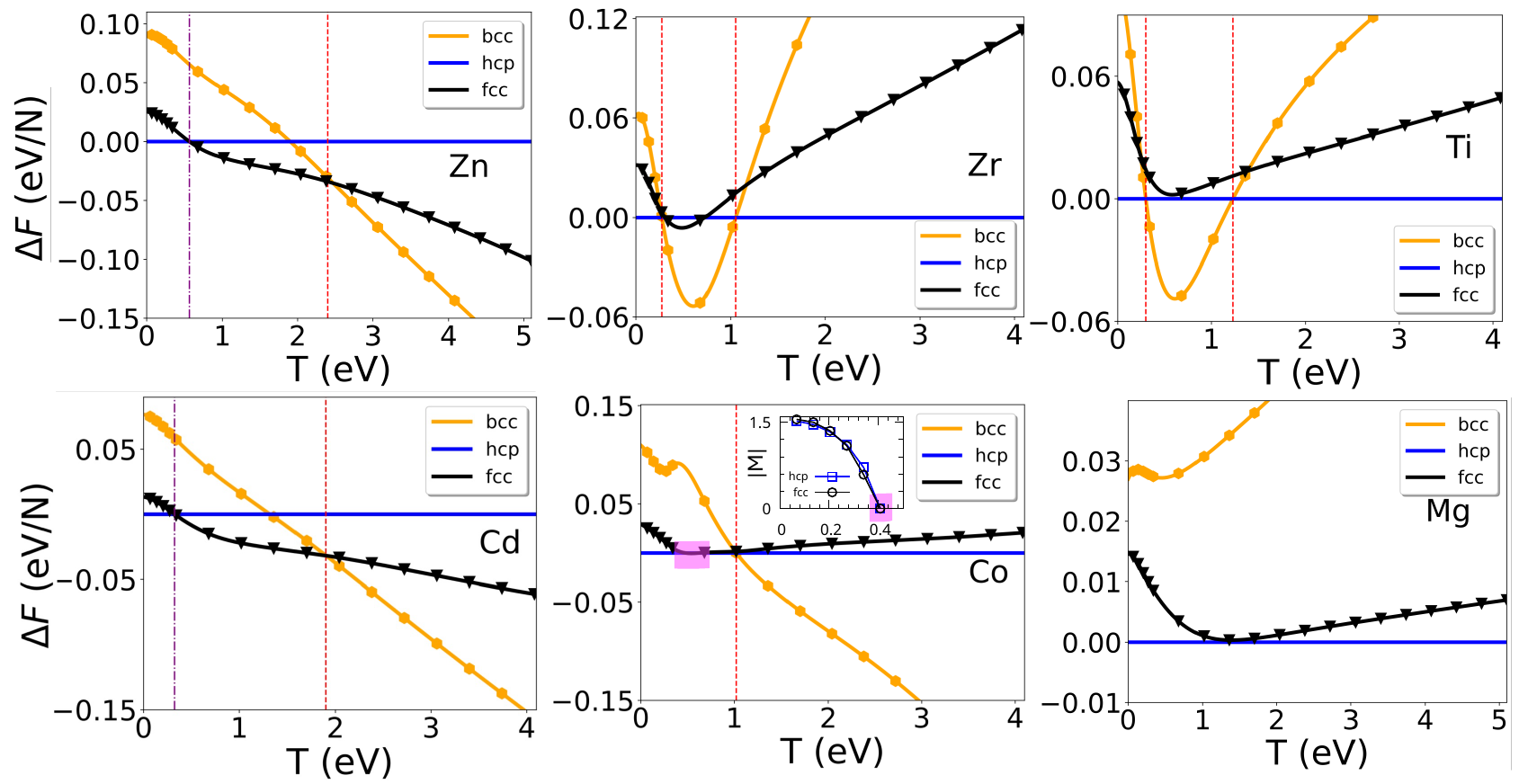}
    \caption{Helmholtz free-energy differences $\Delta F$ between the fcc, bcc, and hcp phases as functions of electronic temperature. Vertical lines indicate the phase transition temperatures. The phase diagram of Co was obtained using spin-polarized DFT and the inset shows the temperature dependence of the absolute magnetization per atom. The vanishing magnetization and the condition $\Delta F<10$ meV/atom occur at the same temperature region, as highlighted.}
    \label{fig:HCP}
\end{figure*}
 
Finite-T DFT based methods have been successfully applied on variety of systems \cite{Driver16,Militzer2021,Wu21,Karasiev16,Karasiev22,SXHu11,Ding18,Bonitz2024}, but less on the transition-metals with complex d-band electronic structure \cite{Lichtenstein01,Katsnelson}. Application of finite-temperature DFT to transition metals has revealed substantial modifications of bonding, magnetism, and lattice dynamics at elevated electronic temperatures \cite{Azadi2024,Azadi2025,Azadi2025II}. Finite-T DFT calculations predict strong hardening of specific phonon branches, changes in effective interatomic forces, and even entropy-driven phase transitions that are absent at equilibrium. These effects are especially pronounced in materials whose DOS exhibits strong energy dependence near the Fermi level, such as 3d and 4d transition metals, where the redistribution of occupation across the d-orbitals produces large electronic-entropy contributions to the free energy. Consequently, finite-temperature DFT has become the primary theoretical tool for describing ultrafast structural dynamics, warm dense matter regimes \cite{Mazevet,Zhang2021}, and phase stability under extreme electronic excitation\cite{Bonitz2020}. In this work, finite-temperature density functional theory is employed to elucidate the role of electronic entropy in reshaping the electronic structure of 17 metals including Zr, Ti, Cd, Zn, Co, and Mg (hcp-group), Ni, Cu, Ag, Al, Pt, and Pb (fcc-group), and Cr, W, V, Nb, and Mo (bcc-group) with hcp, fcc, and bcc structures at ground state. 

\section{Computational details and theoretical background}
The thermodynamic phase diagram was obtained using finite-temperature density functional theory as implemented in the Quantum ESPRESSO package~\cite{QE}. All electronic-structure calculations employed the revised Perdew–Burke–Ernzerhof (PBE) generalized-gradient approximation to the exchange–correlation functional~\cite{PBE,PBEsol}. We considered 17 metals with hcp, fcc, and bcc crystal structures and performed self-consistent calculations for electronic temperatures T up to 7 eV using the Mermin formalism for finite-T DFT. A plane-wave kinetic-energy cutoff of 80 Ry and an augmentation-charge cutoff of 800 Ry were found to provide well-converged total energies and stresses for all temperatures examined. The ionic cores were treated using the PAW tabulated pseudopotential supplied with Quantum ESPRESSO~\cite{QE2}.

Electronic occupations were determined using the standard Fermi–Dirac distribution with temperature T. Since higher electronic temperatures populate states far above the Fermi level, the number of empty bands was increased systematically with T to ensure convergence of the internal energy, entropy, and pressure. All self-consistent finite-T calculations were carried out at fixed cell volume for each crystal structure. For each system, the hcp, fcc, and bcc structures were generated by relaxing the lattice parameters at an electronic temperature of T = 25 meV and zero external pressure, with the crystal symmetry constrained throughout the relaxation. All self-consistent finite-T thermodynamic phase diagrams were obtained at fixed cell volume for each crystal structure. The thermal pressure at each temperature was obtained directly from the electronic stress tensor, which evaluates the derivative $-\partial F/\partial V$ at constant volume.

The temperature dependence of the chemical potential provides the first indication of how electronic excitation differently influences the hcp, fcc and bcc phases of metals. In finite-T DFT, the electronic chemical potential $\mu(T)$ is determined self-consistently from the Fermi–Dirac occupations
$f(\epsilon, T)=\frac{1}{e^{(\epsilon-\mu)/k_{\mathrm{B}}T}+1}$, as the number of electrons is fixed.
At low to moderate electronic temperatures, the variation of $\mu(T)$ can be understood analytically through the Sommerfeld \cite{Ashcroft} expansion. For a transition metal system with density of states $g(\epsilon)$, the chemical potential at finite electronic temperature satisfies
\begin{equation}
 \mu(T)=E_F-\frac{\pi^2}{6}(k_{\mathrm{B}}T)^2
\left.\frac{d\ln g(\epsilon)}{d\epsilon}\right|_{\epsilon=E_F}
+\mathcal{O}(T^4),   
\end{equation}

which shows that the sign and magnitude of the slope $g'(\epsilon)$ at the Fermi level determine whether $\mu(T)$ increases or decreases with temperature. Although hcp, fcc and bcc structure of metals may have similar DOS values at $\epsilon=E_F$, they differ in the slope of the DOS in the vicinity of the Fermi energy. If the Fermi level lies on the descending side of the d-band, $g'(E) < 0$, implying that $\mu(T)$ increases with electronic temperature. A similar behaviour is exactly obtained from finite-T DFT calculations, different phases exhibit a monotonic increase of $\mu(T)$, with the magnitude of the shift reflecting the steepness of the DOS slope near the Fermi level.

The next important contribution is the electronic entropy, which plays a decisive role in determining the relative phase stability under electronic excitation. The electronic entropy is
\begin{equation}
    S(T) = -2k_{\mathrm{B}} \int d\epsilon\, g(\epsilon) \left[f\ln f + (1-f)\ln(1-f)\right],
\end{equation}
and enters the Helmholtz free energy through the $-TS$ term, $F(T) = E(T) - TS(T)$.
Although the DOS values at the Fermi level of hcp, fcc and bcc phases of some of the studied metals are comparable, the shapes of their DOS over the broader energy window $|\epsilon - \mu|\lesssim k_{\mathrm{B}}T$ differ significantly. As electronic temperature increases above the eV range, Fermi–Dirac broadening samples increasingly deep regions of the d-band. Some phases exhibit a flatter and more extended DOS profile in the full thermal window, whereas DOS of other phases exhibit a more rapidly varying structure. Because the entropy kernel $f(1-f)$ weights states symmetrically around $\mu$, the flatter DOS leads to a larger integrated $S(T)$. Consequently, the entropic contribution $-TS$ decreases more rapidly. This entropic advantage is the dominant driving force that lowers the free energy of metallic phases at high electronic temperatures.

\begin{figure*}[!htb]
    \centering
    \includegraphics[width=1.0\linewidth]{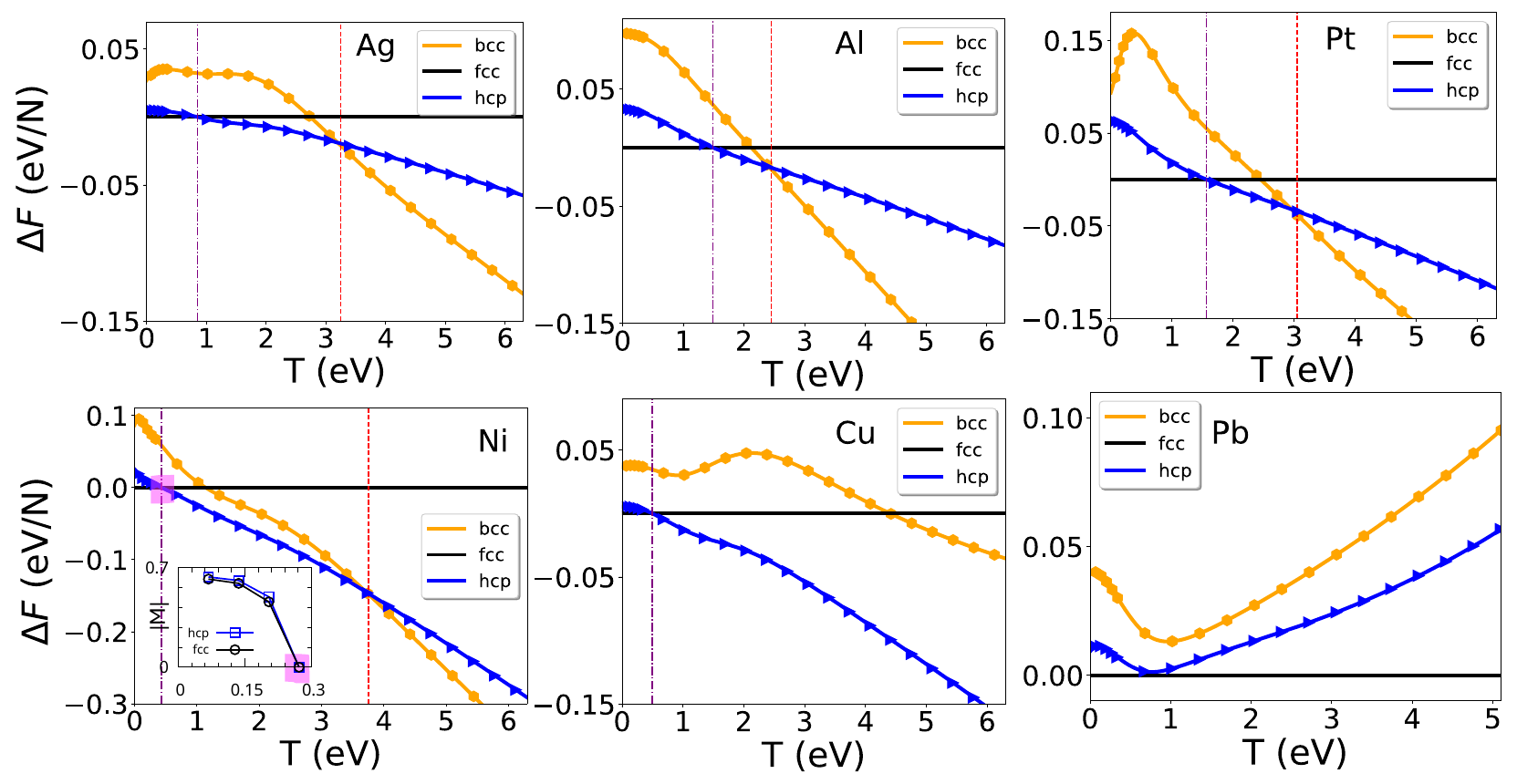}
    \caption{Helmholtz free-energy differences between the bcc, hcp, and fcc phases as functions of electronic temperature. Vertical lines indicate the phase transition temperatures. The phase diagram of Ni was obtained using spin-polarized DFT and the inset shows the temperature dependence of the absolute magnetization per atom. The vanishing magnetization and the condition $\Delta F<10$ meV/atom occur at the same temperature region, as highlighted.}
    \label{fig:FCC}
\end{figure*}

\section{Results and discussion}
\subsection{Electronic-entropy–driven phase stability and density trends in elemental metals}
We first present a systematic analysis of the Helmholtz free-energy phase diagrams and corresponding mass densities for seventeen elemental metals spanning hcp-, fcc-, and bcc-derived ground-state structures. For each element, the free energies of the competing hcp, fcc, and bcc phases were computed as functions of electronic temperature up to T = 7 eV, allowing us to identify electronic-entropy–driven solid–solid phase transitions. In parallel, the equilibrium mass density of each phase was evaluated to elucidate the role of electronic thermal pressure in determining phase stability.

Figure~\ref{fig:HCP} presents the Helmholtz free-energy differences $\Delta F$ between the fcc, bcc, and hcp phases of Zn, Zr, Ti, Cd, Co, and Mg, all of which crystallize in the hcp structure under ambient conditions. For each element, the free energies of the fcc and bcc phases are evaluated relative to the hcp phase as functions of electronic temperature T. In all cases, $\Delta F$ decreases with increasing T, indicating a general tendency toward electronic-entropy–driven solid–solid phase transitions.

At zero temperature, the free-energy difference between the hcp and fcc phases is smaller than that between the hcp and bcc phases for all six elements, suggesting that an hcp-to-fcc transition is generally favored upon electronic excitation. This behavior is indeed observed for Zn and Cd. In contrast, for Zr and Ti the free energy of the bcc phase decreases more rapidly with increasing T than that of the fcc phase, leading to an hcp-to-bcc transition instead. Notably, for Zr the free-energy differences among all three structures are smaller than 5 meV/atom at the transition temperature $T^\prime$, indicating the possibility of polymorphism within a narrow temperature window.

For Co, spin-polarized calculations reveal that the ferromagnetic ordering of both hcp and fcc phases becomes unstable with increasing T, as the absolute magnetic moment per atom vanishes at approximately $T \approx 0.4$ eV. At this temperature, the free-energy difference between the hcp and fcc phases is below 10 meV/atom, highlighting the interplay between electronic entropy and magnetic degrees of freedom. In contrast, no phase transition is predicted for Mg within the investigated temperature range, despite the small free-energy difference between the fcc and hcp phases at T = 1–1.7 eV. 

For Zn and Cd, the fcc structure becomes the thermodynamically stable phase over an intermediate range of electronic temperatures, while the bcc phase is stabilized only at higher electronic temperatures. In contrast, Zr and Ti exhibit qualitatively different behavior as the fcc phase never becomes thermodynamically favorable, and the bcc structure is stabilized in an intermediate temperature range, whereas the hcp phase, which has the lowest free energy at zero temperature, regains stability at higher electronic temperatures. This nonmonotonic sequence reflects the delicate balance between electronic entropy and band structure energetics in early transition metals.

The phase diagram of Co reveals yet another distinct scenario. In this case, the bcc phase becomes stable at relatively low electronic temperatures, $T \sim 1$ eV, coinciding with the loss of ferromagnetic order. This behavior highlights the strong coupling between magnetic degrees of freedom and electronic entropy, which can significantly modify the relative stability of competing crystal structures under electronic excitation \cite{Azadi2024}.

Figure~\ref{fig:FCC} presents the Helmholtz free-energy differences of the bcc and hcp phases relative to the fcc phase for elemental metals that crystallize in the fcc structure at zero temperature. Similar to the hcp-group elements shown in Fig.~\ref{fig:HCP}, the dominant competing phases at zero and low electronic temperatures are fcc and hcp, reflecting their close structural and energetic proximity. With the exception of Pb, all studied elements exhibit an fcc-to-hcp phase transition at relatively low electronic temperatures. In Pb, the free-energy difference between the fcc and hcp phases remains extremely small over the temperature range T $\approx$ 0.7–1 eV, preventing a clear stabilization of the hcp phase.

The phase diagrams of Ag, Ni, and Cu reveal an fcc-to-hcp transition at electronic temperatures below 1 eV. Among these elements, Ni exhibits the lowest transition temperature, which can be attributed to its ferromagnetic ground state. The ferromagnetic ordering in Ni collapses at approximately T $\approx$ 0.3 eV, where the free-energy difference between the fcc and hcp phases becomes minimal, as a result, the solid–solid phase transition occurs just below T $\approx$ 0.5 eV. This behavior highlights the significant role of magnetic entropy in modifying structural stability under electronic excitation.

For Al and Pt, the hcp phase is stabilized only within a relatively narrow temperature window, whereas in Cu the hcp phase remains stable over a broad range of intermediate and high electronic temperatures. At sufficiently high electronic temperatures, most fcc-group elements undergo a transition to the bcc phase, indicating the increasing entropic stabilization of the bcc structure. Notably, Pb constitutes an exception as the fcc phase remains the most favorable structure over the entire electronic temperature range investigated, underscoring its unusual electronic structure robustness against entropy-driven structural reordering. 
\begin{figure*}[!htb]
    \centering
    \includegraphics[width=1.0\linewidth]{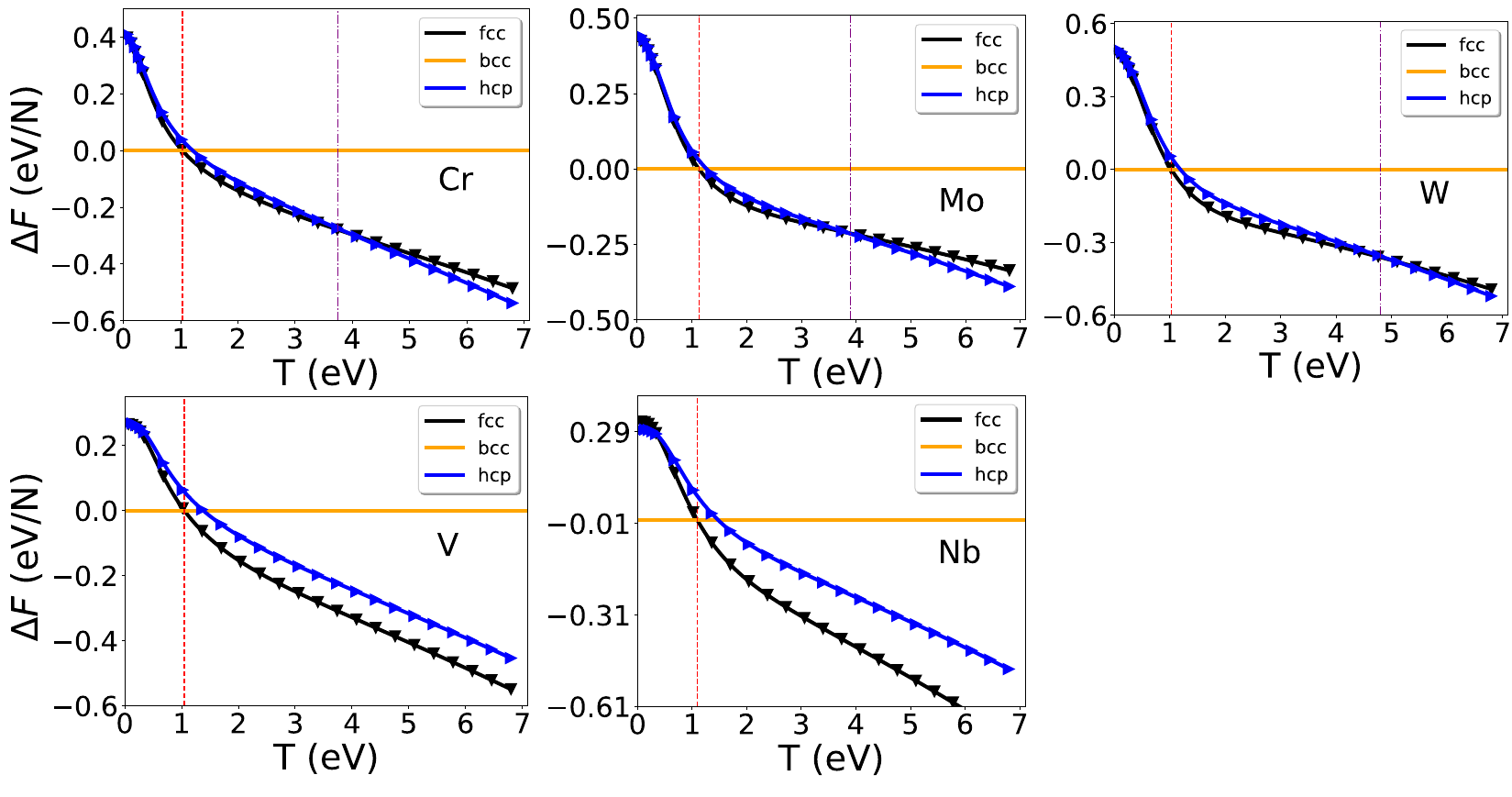}
    \caption{Helmholtz free-energy differences between the fcc, hcp, and bcc phases as functions of electronic temperature. Vertical lines indicate the phase transition temperatures. }
    \label{fig:BCC}
\end{figure*}

The phase diagrams of the bcc-group elements shown in Fig.~\ref{fig:BCC}, which include metals that adopt the bcc structure at zero temperature, display behavior that is qualitatively distinct from that of the fcc- and hcp-group elements. In all bcc-group systems studied, a bcc-to-fcc phase transition occurs at electronic temperatures of approximately T $\sim$ 1 eV, indicating an early destabilization of the bcc structure under electronic excitation.

For Cr, Mo, and W, the free energy difference between the fcc and hcp phases remains small over the entire investigated temperature range, reflecting a strong competition between these two close-packed structures. In these elements, the fcc phase is stabilized over a broad intermediate temperature range, extending from approximately 1 to $\sim$ 4 eV for Cr and Mo, and from 1 to $\sim$ 5 eV for W. At higher electronic temperatures, the hcp phase becomes thermodynamically favorable, completing a sequence of bcc $\rightarrow$ fcc $\rightarrow$ hcp transitions.

In contrast, V and Nb exhibit a markedly different trend. For these elements, the free energy difference between the fcc and hcp phases is significantly larger than in Cr, Mo, and W, suppressing the stabilization of the hcp phase across the entire temperature range considered. As a result, V and Nb undergo only a single bcc-to-fcc phase transition, with the fcc structure remaining stable at higher electronic temperatures.

An important and somewhat counterintuitive observation emerges when comparing the behavior of the bcc phase across different structural groups. While the bcc structure becomes the most favorable phase at high electronic temperatures for many elements in the hcp- and fcc-groups, it is destabilized at relatively low electronic temperatures (T $\sim$ 1 eV) in all bcc-group elements studied and does not reemerge as a stable phase within the investigated temperature range. This contrast highlights the non-universal role of bcc symmetry under electronic excitation and underscores the importance of detailed electronic structure effects in determining entropy-driven phase stability.  

\begin{table}[!htb]
    \centering
    \begin{tabular}{|c|c|c|}
    \hline\hline
    element & 1st transition & 2nd transition \\
    \hline
    Zr & $\text{hcp}\rightarrow \text{bcc}:T^\prime=0.27$& $\text{bcc}\rightarrow \text{hcp}:T^{\prime\prime}=1.05$\\
    Ti & $\text{hcp}\rightarrow \text{bcc}:T^\prime=0.30$& $\text{bcc}\rightarrow \text{hcp}:T^{\prime\prime}=1.22$ \\
    Cd & $\text{hcp}\rightarrow \text{fcc}:T^\prime=0.32$& $\text{fcc}\rightarrow \text{bcc}:T^{\prime\prime}=1.90$ \\
    Zn & $\text{hcp}\rightarrow \text{fcc}:T^\prime=0.57$& $\text{fcc}\rightarrow \text{bcc}:T^{\prime\prime}=2.41$\\    
    Co & $\text{hcp}\rightarrow \text{bcc}:T^\prime=1.02$& \\
    Mg & & \\
    \hline
    Ni &$\text{fcc}\rightarrow \text{hcp}:T^\prime=0.43$& $\text{hcp}\rightarrow \text{bcc}:T^{\prime\prime}=3.75$\\
    Cu &$\text{fcc}\rightarrow \text{hcp}:T^\prime=0.50$& \\
    Ag &$\text{fcc}\rightarrow \text{hcp}:T^\prime=0.85$& $\text{hcp}\rightarrow \text{bcc}:T^{\prime\prime}=3.25$ \\
    Al &$\text{fcc}\rightarrow \text{hcp}:T^\prime=1.49$& $\text{hcp}\rightarrow \text{bcc}:T^{\prime\prime}=2.45$\\
    Pt &$\text{fcc}\rightarrow \text{hcp}:T^\prime=1.56$& $\text{hcp}\rightarrow \text{bcc}:T^{\prime\prime}=3.05$\\
    Pb & & \\
    \hline
    Cr &$\text{bcc}\rightarrow \text{fcc}:T^\prime=1.03$ &$\text{fcc}\rightarrow \text{hcp}:T^{\prime\prime}=3.75$ \\
    W  &$\text{bcc}\rightarrow \text{fcc}:T^\prime=1.03$ &$\text{fcc}\rightarrow \text{hcp}:T^{\prime\prime}=4.8$ \\   
    V  &$\text{bcc}\rightarrow \text{fcc}:T^\prime=1.05$ & \\
    Nb &$\text{bcc}\rightarrow \text{fcc}:T^\prime=1.10$ & \\    
    Mo &$\text{bcc}\rightarrow \text{fcc}:T^\prime=1.13$ &$\text{fcc}\rightarrow \text{hcp}:T^{\prime\prime}=3.90$ \\
    \hline\hline
    \end{tabular}
    \caption{First and second phase transition temperature for all the studied metallic systems.}
    \label{tab:T_trans}
\end{table}

\begin{figure*}
    \centering
    \includegraphics[width=0.95\linewidth]{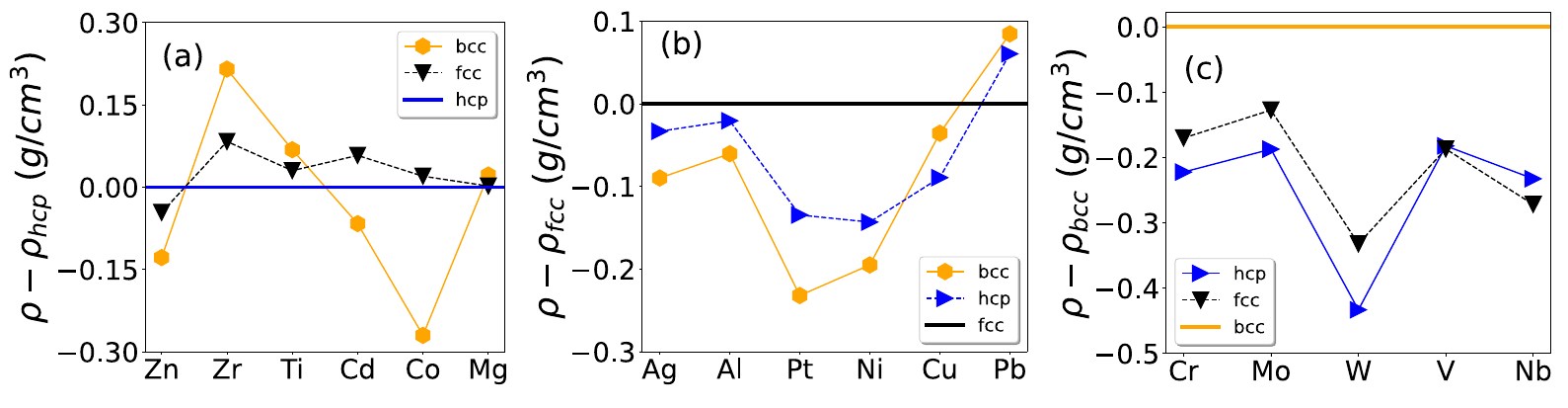}
    \caption{(a) The behavior of density of fcc and bcc phases with respect to hcp for hcp-group elements. (b) The behavior of density of hcp and bcc phases with respect to fcc for fcc-group elements. (c) The behavior of density of hcp and fcc phases with respect to bcc for bcc-group elements.}
    \label{fig:density}
\end{figure*}

Analysis of the DFT-predicted mass densities $\rho$ at ground state for the bcc-group elements reveals systematic trends among the competing crystal structures. For Cr, Mo, and W, the densities follow the ordering $\rho_{\mathrm{bcc}} > \rho_{\mathrm{fcc}} > \rho_{\mathrm{hcp}}$, whereas for V and Nb the ordering is $\rho_{\mathrm{bcc}} > \rho_{\mathrm{hcp}} > \rho_{\mathrm{fcc}}$ (Fig.~\ref{fig:density}). When these density trends are considered together with the calculated phase diagrams, a common behavior emerges indicating all bcc-group elements undergo transitions toward crystal structures with lower density as the electronic temperature increases. This observation indicates that electronic-entropy–driven phase transitions in bcc metals are accompanied by an effective volume expansion, consistent with the destabilization of densely packed electronic states under strong electronic excitation.

The systematic tendency of bcc-group elements to transform toward lower-density crystal structures with increasing electronic temperature can be naturally interpreted in terms of hot-electron thermal pressure \cite{Azadi2025}. As the electronic temperature rises, the electronic contribution to the pressure increases due to the thermal population of higher-energy electronic states, effectively generating an expansive force on the lattice even in the absence of ionic heating. In bcc transition metals, the ground-state stability of the bcc phase is closely tied to relatively compact electronic charge distributions and band-structure features that minimize the electronic pressure at low temperatures. Thermal smearing of the Fermi surface at elevated electronic temperatures weakens these stabilizing features, increasing the electronic pressure and favoring structures with larger equilibrium volumes. This mechanism is consistent with the observed sequence of bcc$\rightarrow$fcc and bcc$\rightarrow$hcp transitions, both of which involve a reduction in mass density relative to the bcc phase. The close correspondence between density trends and phase stability across the bcc-group therefore supports a picture in which electronic thermal pressure, rather than lattice entropy or ionic motion, acts as a primary driving force for entropy-induced structural transformations under strong electronic excitation.

A similar tendency toward reduced mass density at high electronic excitation is also observed for the hcp-group elements. For Zn, Cd, and Co, the bcc phase possesses a lower density than the hcp phase, which promotes an hcp-to-bcc transition at intermediate electronic temperatures as electronic entropy and thermal pressure increase. This effect is particularly pronounced in Co, where the density of the bcc phase is substantially lower than that of the hcp phase, consistent with the stabilization of the bcc structure at relatively low electronic temperatures of order T $\sim$ 1 eV. In contrast, the hcp phase of Mg remains thermodynamically stable over the entire temperature range investigated, which can be directly related to the fact that the hcp structure in Mg has a lower density than both the fcc and bcc phases. These observations reinforce the broader conclusion that electronic-entropy–driven phase transitions are closely linked to density reduction and electronic thermal pressure, providing a unifying physical picture across different structural groups.

The density trends observed for the fcc-group elements further demonstrate the close connection between density reduction and phase stability under increasing electronic entropy. For all fcc-group elements except Pb, the bcc phase exhibits the lowest mass density at high electronic temperatures, providing a natural driving force for the stabilization of the bcc structure as electronic thermal pressure increases. Consequently, these elements undergo a transition to the bcc phase at sufficiently high electronic temperatures. Pb constitutes a notable exception to this trend as in this case, the fcc phase remains the lowest-density structure across the entire temperature range studied, which explains its persistent stability and the absence of any solid–solid phase transition. These results reinforce the view that electronic-entropy–driven phase transitions in fcc metals are governed by a competition between electronic thermal pressure and the density hierarchy of competing crystal structures.

Taken together, the phase diagrams of the hcp-, fcc-, and bcc-group elements reveal both universal features and pronounced group-specific behavior under electronic excitation. A common trend across all three groups is the strong reduction of free-energy differences between competing crystal structures with increasing electronic temperature, underscoring the central role of electronic entropy in reshaping structural stability. In this sense, electronic excitation acts as a unifying thermodynamic control parameter that can induce solid–solid phase transitions even in elemental metals with simple crystal structures.

Despite this universality, the sequence of stabilized phases and the fate of the ground state structure differ markedly among the three groups, as summarized in Tab.~\ref{tab:T_trans}. In the hcp- and fcc-derived elements, the bcc structure is frequently stabilized at high electronic temperatures, reflecting its large electronic entropy and broad d-band characteristics. In contrast, for bcc-derived elements the bcc phase is destabilized at relatively low electronic temperatures and never reemerges as the most favorable structure within the studied range. This asymmetry highlights that high-entropy stabilization of bcc is not intrinsic to the structure itself, but depends sensitively on band filling and the detailed electronic DOS of each element.

Another important distinction lies in the competition between fcc and hcp phases. For hcp- and fcc-derived elements, these two close-packed structures dominate the low- and intermediate-temperature regimes, often separated by only a few meV/atom, enabling multiple phase transitions or narrow polymorphic windows. In the bcc group, however, the fcc–hcp competition becomes relevant only after the bcc phase is destabilized, with the relative stability of fcc and hcp governed by subtle differences in electronic entropy at higher temperatures. These comparisons emphasize that while electronic entropy universally lowers structural energy scales, the resulting phase sequence is dictated by element-specific electronic structure rather than crystal symmetry alone.

\subsection{Case study: anomalous behavior of Zr}
Zr provides a particularly instructive case study because its electronic-entropy–driven phase behavior deviates from the general density-based trends observed in most hcp-group elements. While the stabilization of high-temperature phases in hcp- and fcc-derived elements can typically be rationalized by a tendency toward lower-density structures under increasing electronic thermal pressure, Zr exhibits a qualitatively different response. Specifically, despite the bcc phase having a higher mass density than both the hcp and fcc phases, it becomes thermodynamically stable over a relatively narrow intermediate electronic temperature range, 0.27 $<$ T $<$ 1.01 eV. This behavior cannot be explained solely in terms of density reduction. A similar trend is also observed in Ti, however, the effect is more pronounced in Zr due to the larger density difference between the hcp and bcc phases. In the following, we focus on Zr to elucidate the microscopic origin of this density-opposing stabilization of the bcc phase, with particular emphasis on the electronic density of states.
\begin{figure*}
    \centering
    \includegraphics[width=1.0\linewidth]{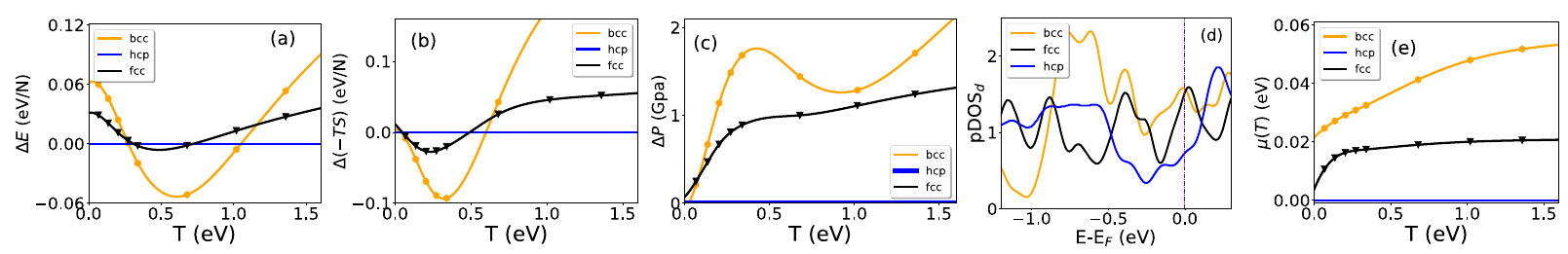}\\
    \caption{The internal energy E (a), and electronic entropy term -TS (b) of fcc and bcc phases of Zr with respect to hcp.  (c) Thermal pressure of fcc and bcc phases of Zr with respect to hcp versus temperature. (d) Projected electronic density of states of d-band (pDOS$_d$) of hcp, fcc and bcc phases near the Fermi energy. The vertical line shows the Fermi energy. (e) Chemical potential $\mu(T)$ of of fcc and bcc phases of Zr with respect to hcp as a function of T.}
    \label{fig:Zr_plot}
\end{figure*}

In Zr, the hcp-to-bcc phase transition occurs at a relatively low electronic temperature of T$^\prime$ $\approx$ 0.27 eV. The bcc phase is stabilized only within a narrow intermediate temperature window, spanning approximately 0.27 $\leq$ T $\leq$ 1.05 eV, above which the system reverts to the hcp structure. This behavior motivates a focused analysis of the electronic states within an energy window of approximately 1 eV below the Fermi level, where thermal occupation effects are expected to be most pronounced.

A decomposition of the Helmholtz free energy reveals that both energetic and entropic contributions play essential roles in stabilizing the bcc phase (Fig.~\ref{fig:Zr_plot}). Comparison of the internal energies shows that the internal energy of the bcc phase drops below that of the hcp phase at the transition temperature T$^\prime\approx$ 0.27 eV, indicating an electronic-energy–driven preference for the bcc structure. Simultaneously, the entropic contribution, -TS, exhibits a minimum for the bcc phase that is lower than that of the hcp phase at the same temperature, suggesting that electronic entropy further reinforces the stabilization of the bcc structure.

In addition, the electronic thermal pressure provides a complementary perspective on the transition (Fig.~\ref{fig:Zr_plot}). The thermal pressure associated with hot electrons in the bcc phase reaches a maximum relative to the hcp phase near the transition temperature, reflecting enhanced sensitivity of the bcc electronic states to thermal excitation. Together, these results show that in Zr the stabilization of the bcc phase arises from a cooperative interplay between electronic energy lowering, enhanced electronic entropy, and hot-electron thermal pressure, which collectively overcome the density-based tendency that would otherwise favor the hcp structure.

The DOS in the vicinity of the Fermi energy in Zr is dominated by d-band contributions, making the projected d-band DOS ($\mathrm{pDOS}_d$) particularly relevant for understanding the phase transition. Near the hcp-to-bcc transition temperature, the $\mathrm{pDOS}_d$ of the hcp phase exhibits a minimum below the Fermi level, whereas the $\mathrm{pDOS}_d$ of the bcc phase is more than two times larger at the same energy (Fig.~\ref{fig:Zr_plot}). This substantial enhancement of the d-band DOS in the bcc phase leads to a significantly larger electronic entropy contribution, which favors the stabilization of the bcc structure despite its higher mass density.

In addition to the magnitude of the DOS, the energy dependence of the $\mathrm{pDOS}_d$ near the Fermi level differs between the two phases. The contrasting slopes of the DOS influence the temperature dependence of the chemical potential, which can be described by the Sommerfeld expansion \cite{Ashcroft}, causing the chemical potential of the bcc phase to increase more rapidly with electronic temperature than that of the hcp phase (Fig.~\ref{fig:Zr_plot}). This difference further lowers the electronic free energy of the bcc phase at intermediate temperatures by enhancing thermal occupation asymmetry around the Fermi level. Together, the larger d-band DOS and its distinct energy slope provide a microscopic explanation for the electronic-energy and entropy-driven stabilization of the bcc phase in Zr at intermediate electronic temperatures.

A clear influence of the d-band electronic structure on phase stability is also evident at energies approximately 1 eV below the Fermi level. In this energy range, the $\mathrm{pDOS}_d$ of the bcc phase is much smaller than that of the hcp phase. Notably, the $\mathrm{pDOS}_d$ of the bcc phase nearly vanishes at the electronic temperature where the system undergoes the bcc-to-hcp transition. This depletion of d-band states around the phase transition temperature reduces the electronic entropy gain and energetic stabilization of the bcc phase, thereby destabilizing it relative to the hcp structure. These observations highlight the crucial role of the detailed energy dependence of the d-band DOS in governing the nonmonotonic phase stability of Zr under electronic excitation.

\section{Summary and Outlook}
In summary, the present results establish electronic entropy as a dominant factor governing structural stability in elemental metals under strong electronic excitation. Across hcp-, fcc-, and bcc-derived systems, finite-temperature free-energy calculations reveal that modest increases in electronic temperature of order 1–5 eV are sufficient to induce solid–solid phase transitions between competing crystal structures. These transitions occur in the absence of ionic heating and are driven by entropy-induced reshaping of the electronic free-energy landscape, particularly through changes in the density of states near the Fermi level and the thermal smearing of band-structure features that stabilize the ground-state phase. The resulting phase sequences and transition temperatures depend sensitively on band filling, magnetic order, and the relative DOS of competing structures, leading to element-specific behavior despite broadly similar crystal symmetries.

Across the hcp-, fcc-, and bcc-group elements, a consistent relationship emerges between electronic excitation, density reduction, and structural stability. In all three groups, increasing electronic temperature enhances the electronic thermal pressure, favoring crystal structures with lower mass density and larger equilibrium volumes per atom. For hcp-group elements, this mechanism stabilizes the bcc phase when it possesses a lower density than the hcp structure, while elements such as Mg remain hcp-stable because the hcp phase is already the least dense. In the fcc group, the bcc phase generally becomes the lowest-density structure at high electronic temperatures, leading to a ubiquitous fcc-to-bcc transition, with Pb as a notable exception where the fcc phase remains the least dense and thus stable across the entire temperature range. In contrast, for bcc-group elements, electronic excitation destabilizes the initially dense bcc phase at relatively low electronic temperatures, driving transitions toward less dense fcc or hcp structures without re-stabilizing bcc at higher temperatures. Taken together, these trends demonstrate that electronic thermal pressure provides a unifying physical mechanism for entropy-driven solid–solid phase transitions in elemental metals, with density reduction serving as a key indicator of structural stabilization under strong electronic excitation.

From an experimental perspective, the predicted transitions are directly relevant to ultrafast pump–probe studies of laser-excited metals, where electronic temperatures of several electronvolts are routinely achieved on sub-100 fs timescales while the lattice remains near its initial temperature. The electronic-entropy–driven phase transitions identified here are expected to manifest as transient symmetry changes, phonon hardening \cite{Azadi2025,Azadi2024}, or the emergence of new Bragg features in time-resolved X-ray or electron diffraction measurements, preceding conventional thermal effects such as lattice expansion or melting. In magnetic systems such as Co and Ni, the coupling between magnetic collapse and structural instability provides an additional experimental signature, linking ultrafast demagnetization to lattice reorganization. More broadly, these results suggest that femtosecond laser excitation can be used as a selective and reversible tool to access metastable crystal structures in elemental metals, opening new pathways for ultrafast structural control and non-equilibrium phase engineering.

Our work is computational and predicts electronic-entropy–driven modifications of the electronic free-energy landscape that can modify, and perhaps even stabilize, crystal structures on ultrafast electronic timescales. Experimental studies have established a broad range of non-equilibrium phenomena in metals \cite{Vinko12}, including nonthermal melting \cite{Rousse}, ultrafast magnetic and structural switching, and transient lattice instabilities \cite{Kraus25}, but these effects have generally been interpreted in regimes where electronic and lattice excitations are strongly coupled. Ultrafast X-ray techniques at free-electron laser facilities have further enabled the observation of non-equilibrium phase transitions in selected materials. However, most demonstrations to date have focused on shock-compressed systems, in which mechanical, thermal, and electronic contributions cannot be readily disentangled \cite{Lindenberg00,Milathianaki13,Lindroth19}.

The regime addressed here, where electronic entropy alone reshapes the free-energy landscape and drives a solid–solid structural rearrangement on femtosecond-to-picosecond timescales while the lattice remains comparatively cold, has not yet been observed experimentally. Recent advances in pump–probe capabilities at X-ray free-electron laser facilities provide a realistic opportunity to access this regime by combining strong electronic excitation with ultrafast time-resolved X-ray diffraction probes. Such experiments would directly test whether the entropic contributions to the electronic free energy identified here govern phase stability under extreme non-equilibrium conditions, while simultaneously providing a stringent benchmark for finite-temperature density functional theory for highly-excited systems in the ultrafast regime.

S.A, A.P, and M.S.B acknowledge the support of the Leverhulme Trust under the grant agreement RPG-2023-253. S. Azadi and T.D. K\"{u}hne acknowledge the computing time provided to them on the high-performance computers Noctua2 at the NHR Center in Paderborn (PC2).
S.M.V. acknowledges support from the UK EPSRC under grant EP/W010097/1.
The data that support the findings of this article are openly available \cite{github}, embargo periods may apply.

\bibliography{main}

\end{document}